\documentclass[draft]{agujournal2019}
\usepackage{url} %this package should fix any errors with URLs in refs.
\usepackage{lineno}
\usepackage[inline]{trackchanges} %for better track changes. finalnew option will compile document with changes incorporated.
\usepackage{soul}
\draftfalse

\journalname{AGU Advances}

\begin{document}

%%%%%%%%%%%%%%%%%%%%%%%%%%%%%%%%%%%%%%%%%%%%%%%
%  TITLE
%
% (A title should be specific, informative, and brief. Use
% abbreviations only if they are defined in the abstract. Titles that
% start with general keywords then specific terms are optimized in
% searches)
%
%%%%%%%%%%%%%%%%%%%%%%%%%%%%%%%%%%%%%%%%%%%%%%%

% Example: \title{This is a test title}

\title{Dynamical Tests of a Deep-Learning Weather Prediction Model}

%%%%%%%%%%%%%%%%%%%%%%%%%%%%%%%%%%%%%%%%%%%%%%%
%
%  AUTHORS AND AFFILIATIONS
%
%%%%%%%%%%%%%%%%%%%%%%%%%%%%%%%%%%%%%%%%%%%%%%%

% Authors are individuals who have significantly contributed to the
% research and preparation of the article. Group authors are allowed, if
% each author in the group is separately identified in an appendix.)

% List authors by first name or initial followed by last name and
% separated by commas. Use \affil{} to number affiliations, and
% \thanks{} for author notes.
% Additional author notes should be indicated with \thanks{} (for
% example, for current addresses).

% Example: \authors{A. B. Author\affil{1}\thanks{Current address, Antartica}, B. C. Author\affil{2,3}, and D. E.
% Author\affil{3,4}\thanks{Also funded by Monsanto.}}

\authors{Gregory J. Hakim\affil{1} and Sanjit Masanam\affil{2}}

\affiliation{1}{Department of Atmospheric Sciences, University of Washington, Seattle, WA}
\affiliation{2}{Department of Physics, University of California at Santa Barbara, Santa Barbara, CA}
% \affiliation{3}{Third Affiliation}
% \affiliation{4}{Fourth Affiliation}

%\affiliation{=number=}{=Affiliation Address=}
%(repeat as many times as is necessary)

% Corresponding author mailing address and e-mail address:

% (include name and email addresses of the corresponding author.  More
% than one corresponding author is allowed in this LaTeX file and for
% publication; but only one corresponding author is allowed in our
% editorial system.)

% Example: \correspondingauthor{First and Last Name}{email@address.edu}

\correspondingauthor{Gregory J. Hakim}{ghakim@uw.edu}

%%%%%%%%%%%%%%%%%%%%%%%%%%%%%%%%%%%%%%%%%%%%%%%
% KEY POINTS
%%%%%%%%%%%%%%%%%%%%%%%%%%%%%%%%%%%%%%%%%%%%%%%
%  List up to three key points (at least one is required)
%  Key Points summarize the main points and conclusions of the article
%  Each must be 140 characters or fewer with no special characters or punctuation and must be complete sentences

% Example:
 \begin{keypoints}
 \item The Pangu-weather deep-learning weather prediction model exhibits physically realistic dynamical behavior
 \item Steady tropical heating produces a Matsuno--Gill response in the tropics, and planetary waves that radiate into the extratropics
 \item Localized initial conditions produce realistic hurricanes, extratropical cyclones, and adjustment to geostrophic balance
 \end{keypoints}

%%%%%%%%%%%%%%%%%%%%%%%%%%%%%%%%%%%%%%%%%%%%%%%
%
%  ABSTRACT and PLAIN LANGUAGE SUMMARY
%
% A good Abstract will begin with a short description of the problem
% being addressed, briefly describe the new data or analyses, then
% briefly states the main conclusion(s) and how they are supported and
% uncertainties.

% The Plain Language Summary should be written for a broad audience,
% including journalists and the science-interested public, that will not have 
% a background in your field.
%
% A Plain Language Summary is required in GRL, JGR: Planets, JGR: Biogeosciences,
% JGR: Oceans, G-Cubed, Reviews of Geophysics, and JAMES.
% see http://sharingscience.agu.org/creating-plain-language-summary/)
%
%%%%%%%%%%%%%%%%%%%%%%%%%%%%%%%%%%%%%%%%%%%%%%%

%% \begin{abstract} starts the second page

 \begin{abstract} % ----- 250words -----
   
   Global deep-learning weather prediction models have recently been shown to produce forecasts that rival those from physics-based models run at operational centers. It is unclear whether these models have encoded atmospheric dynamics, or simply pattern matching that produces the smallest forecast error. Answering this question is crucial to establishing the utility of these models as tools for basic science. Here we subject one such model, Pangu-weather, to a set of four classical dynamical experiments that do not resemble the model training data. Localized perturbations to the model output and the initial conditions are added to steady time-averaged conditions, to assess the propagation speed and structural evolution of signals away from the local source. Perturbing the model physics by adding a steady tropical heat source results in a classical Matsuno--Gill response near the heating, and planetary waves that radiate into the extratropics. A localized disturbance on the winter-averaged North Pacific jet stream produces realistic extratropical cyclones and fronts, including the spontaneous emergence of polar lows. Perturbing the 500hPa height field alone yields adjustment from a state of rest to one of wind--pressure balance over $\sim$6 hours. Localized subtropical low pressure systems produce Atlantic hurricanes, provided the initial amplitude exceeds about 5 hPa, and setting the initial humidity to zero eliminates hurricane development. We conclude that the model encodes realistic physics in all experiments, and suggest it can be used as a tool for rapidly testing ideas before using expensive physics-based models.
   
\end{abstract}

\section*{Plain Language Summary} % ----- 200words -----
%Enter your Plain Language Summary here or delete this section.
%Here are instructions on writing a Plain Language Summary: 
%https://www.agu.org/Share-and-Advocate/Share/Community/Plain-language-summary
Deep-learning weather forecast models have recently been shown to be as skillful as the best physics-based models, but it is unclear if they have encoded the laws of physics or simply an effective pattern-matching algorithm. Here we test one such model, Pangu-weather, with four idealized experiments aimed at assessing the physical realism of the model solution: steady tropical heating, extratropical cyclone development, geostrophic adjustment, and hurricane development. A common aspect of these experiments is perturbations that are local in space, since signal propagation from a local source is limited by physics. The outcomes of these four experiments are well-known from observations and theory, so they provide a useful basis for such an evaluation. In all cases, Pangu-weather produces physically realistic solutions that are qualitatively, if not quantitatively, consistent with the known outcomes for these experiments. This suggests that the model has encoded physical constraints that are applicable outside the realm of the data the model was trained on. We suggest an exciting new approach to science, where many hypotheses are rapidly tested using a deep-learning model, and a smaller set of promising results tested using physics-based models.

%%%%%%%%%%%%%%%%%%%%%%%%%%%%%%%%%%%%%%%%%%%%%%%
%
%  BODY TEXT
%'
%%%%%%%%%%%%%%%%%%%%%%%%%%%%%%%%%%%%%%%%%%%%%%%

\section{Introduction}

In the past few years, deep-learning (DL) weather prediction models demonstrate forecast skill comparable to those from government operational centers \cite{weyn2019,bi2023,kurth2023,lam2022}. These models are trained on ERA5 analyses and have forecast skill on initial conditions independent of their training data. In contrast to DL approaches that explicitly enforce physical constraint \cite<e.g.,>{beucler2021}, it is unclear whether these models have encoded atmospheric physics, such as the dynamics of air motion and propagation of disturbances, or simply patterns that minimize the squared error of the next pattern in a sequence.  Physical tests that examine the evolution of spatially localized disturbances are particularly effective in analyzing model physics, since the propagation of signals away from these disturbances is constrained by dynamics. For example, in the small-amplitude limit, the group velocity in linear wave theory sets the speed of energy dispersion away from a local disturbance.

Here we apply the localized-disturbance approach to the Pangu-weather model of \citeA{bi2023} using four canonical experiments, one involving perturbations to the model output, and the other three to perturbed initial conditions. The perturbations are applied to climatological time-mean steady states, which are smoother than any individual state that the model was trained on. These experiments are subjectively chosen, and solutions are not compared directly to identical experiments in a physics-based model, but provide an important plausibility study to motivate such additional experiments. Our hypothesis at the start of this research was that localized features will immediately produce a global response, because no constraint was imposed to prevent this during model training.

In addition to running orders of magnitude faster than physics-based models, these experiments with the Pangu-weather model are comparatively easy to configure. Performing any one of the experiments described here with a modern physics-based weather model is a significant undertaking, primarily due to complexities associated with model initialization. Consequently, if these models can be shown to produce physically realistic solutions, they offer an enormous opportunity for hypothesis testing much faster than is currently possible.

We proceed in section (\ref{sec:method}) with a description of the experiments and the data used to conduct them. Results for the four experiments described above are presented in section (\ref{sec:results}). Conclusions are drawn in section (\ref{sec:conclusions}).

\section{Method and experiment design\label{sec:method}}

The Pangu-weather model uses a vision-transformer architecture trained on ERA5 reanalysis data from 1979--2017 \cite{bi2023}; the trained model weights are publicly available. Model variables consist of global gridded fields of geopotential height, specific humidity of water vapor, temperature, and vector wind components on 13 isobaric levels (1000, 925, 850, 700, 600, 500, 400, 300, 250, 200, 150, 100, and 50hPa), and surface fields (mean-sea-level pressure, 2m air temperature, and 10m vector wind components). Data reside on the native 0.25$^{\circ}$ degree latitude--longitude grid of ERA5. There are four models, which are trained separately: 1h, 3h, 6h, and 24h. \citeA{bi2023} indicate that solutions are most accurate when using the sequence of models with the largest possible time steps to reach a desired lead time (e.g., a 32 hour forecast uses the 24h model, followed by the 6h model and then two steps of the 1h model).

Our experiments involve adding perturbations to a steady climatological-mean atmosphere. We perform the simulations by solving
\begin{equation}
  {\bf x}(t+1) = {\bf N}({\bf x}(t)) - {\bf d\overline{x}}+ {\bf f}.
  \label{eqn:model}
\end{equation}
\noindent Here, ${\bf x}$ represents the model state vector, ${\bf N}$ the Pangu-weather model, and $t$ time indexed according to the version of the model (i.e., $t+1$ means a one-day forecast when using the 24-hour version of the model, and a 3-hour forecast for the 3-hour version). ${\bf f}$ is a modification to the model output, taken here to be zero for all experiments except steady tropical heating, when it is fixed at a specified value. ${\bf d\overline{x}}$ represents the one-step solution of the model that renders the climatological mean atmosphere steady state:
\begin{equation}
  {\bf d\overline{x}} = \bf{N}({\bf \overline{x}}) - {\bf \overline{x}}.
  \label{eqn:mean_state}
\end{equation}
\noindent We may then take ${\bf \overline{x}}$ independent of time. The full state vector, which we send to the Pangu-weather model, is defined by ${\bf x} =  {\bf \overline{x}} + {\bf x}^{\prime}$ where ${\bf x}^{\prime}$ are anomalies from the climatological mean state. For ${\bf x}^{\prime} = {\bf f} = 0$, (\ref{eqn:model}) with (\ref{eqn:mean_state}) gives ${\bf \overline{x}}(t+1) = \bf{N}({\bf \overline{x}}) - \bf{N}({\bf \overline{x}}) +  {\bf \overline{x}} =  {\bf \overline{x}}$; i.e., $ {\bf \overline{x}}$ is time independent.

Taking a leading-order Taylor approximation to ${\bf N}({\bf x}(t))$, and using (\ref{eqn:mean_state}) in (\ref{eqn:model}), gives a conceptual model for the perturbations,
  \begin{equation}
{\bf x}(t+1) ^{\prime} \approx {\bf N}^{\prime} ({\bf x}^{\prime} (t)) + {\bf f},
\end{equation}
\noindent where ${\bf N}^{\prime}$ is the gradient of ${\bf N}$ with respect to ${\bf x}$ evaluated at ${\bf \overline{x}}$. We emphasize that we actually solve (\ref{eqn:model}) and compute ${\bf x}^{\prime} = {\bf x} - {\bf \overline{x}}$ from the solution for all $t$; in other words, the solutions are fully nonlinear.

Since we are interested in spatially localized perturbations, we define ${\bf f}$ and the initial perturbations ${\bf x}^{\prime}(t=0)$ using a function that decays to zero from a local maximum at a specified distance. For this purpose we use the function defined by \citeA[hereafter, GC]{gaspari1999} and define the distance at which the disturbance reaches zero by $L$. Specifics of the four experiments follow.

For the steady heating experiment, we set ${\bf f}$ to be a constant vector with zeros everywhere except for the temperature field within a horizontal region at all levels between 1000hPa and 200 hPa, where it is set to 0.1K(day)$^{-1}$. The region is defined in longitude by the GC function with $L=10,000$km and centered at 120$^{\circ}$E, and in latitude, $\phi$, by $cos(6\phi)$ within 15$^{\circ}$ of the Equator. The initial condition is given by ${\bf \overline{x}}$, which is set to the December-January-February (DJF) ERA5 time average.

For the extratropical cyclone experiment, we define the anomaly field ${\bf x}^{\prime}(t=0)$  by regressing all fields in the state vector against a standardized time series of DJF 500hPa geopotential height at the point 40$^{\circ}$N, 150$^{\circ}$E. The regressed field is then multiplied by the GC function with $L=2000$km to insure a spatially localized disturbance, and added to the DJF time-mean field. We use the same perturbation initial condition for the geostrophic adjustment experiment, except we set to zero all variables at all levels except the 500hPa geopotential height.

For the hurricane experiments we take the same approach as for the extratropical cyclone experiment, except in this case we use the July-August-September (JAS) mean state. The disturbance is defined by regressing all fields in the state vector against a standardized time series of JAS mean-sea-level pressure at the point 15$^{\circ}$N, 40$^{\circ}$W. The regressed field is then multiplied by the GC function with $L=1000$km and added to the JAS time-mean field. For the results in this case we perform simulations by scaling the perturbation field by a multiplicative constant to vary the strength of the initial low-pressure system.

\section{Results\label{sec:results}}

\subsection{Steady tropical heating\label{subsec:heating}}

\begin{figure}[h!]
\noindent\includegraphics[height=.9\textheight]{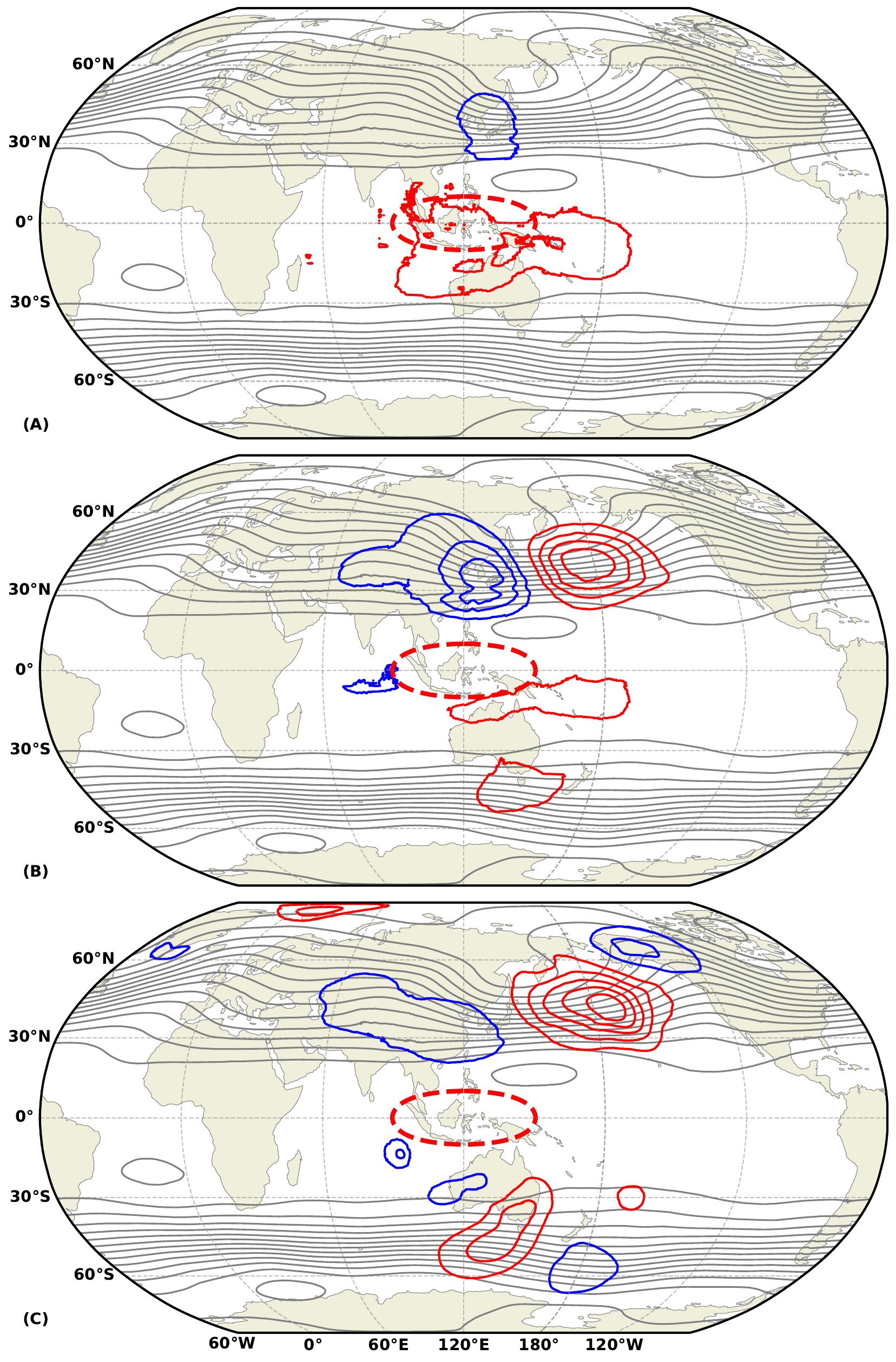}
\caption{Response in DJF 500 hPa geopotential height to steady tropical heating of 0.1 K day$^{-1}$ within the region outlined by the dashed red line. The DJF-averaged geopotential height is shown by gray lines every 60m, and anomalies by red (positive) and blue (negative) lines; the zero contour is suppressed. Solutions are shown for (A) 5 days (contours every 0.3m); (B) 10 days (contours every 2m); (C) 20 days (contours every 20m).\label{fig:heating_500}}
\end{figure}

The Pangu-weather response to weak DJF tropical heating (0.1 K/day), shows a small 500 hPa height increase over the heating region after 5 days, with a negative anomaly to the north (Fig.~\ref{fig:heating_500}A). The extratropical wave train extends downstream and increases in amplitude during days 5--20, with maximum anomalies over 100m at day 20 (Fig.~\ref{fig:heating_500}B,C). A wave-train appears in both hemispheres, with larger amplitude in the northern (wintertime) hemisphere, which has the stronger waveguide for stationary waves. This response is qualitatively similar to classical results \cite<e.g.,>[]{hoskins1981,sardeshmukh1988}, with differences in details dependent on the location, shape, and temporal structure of the heating, seasonality, and other factors.

\begin{figure}[h]
\noindent\includegraphics[width=.9\textwidth]{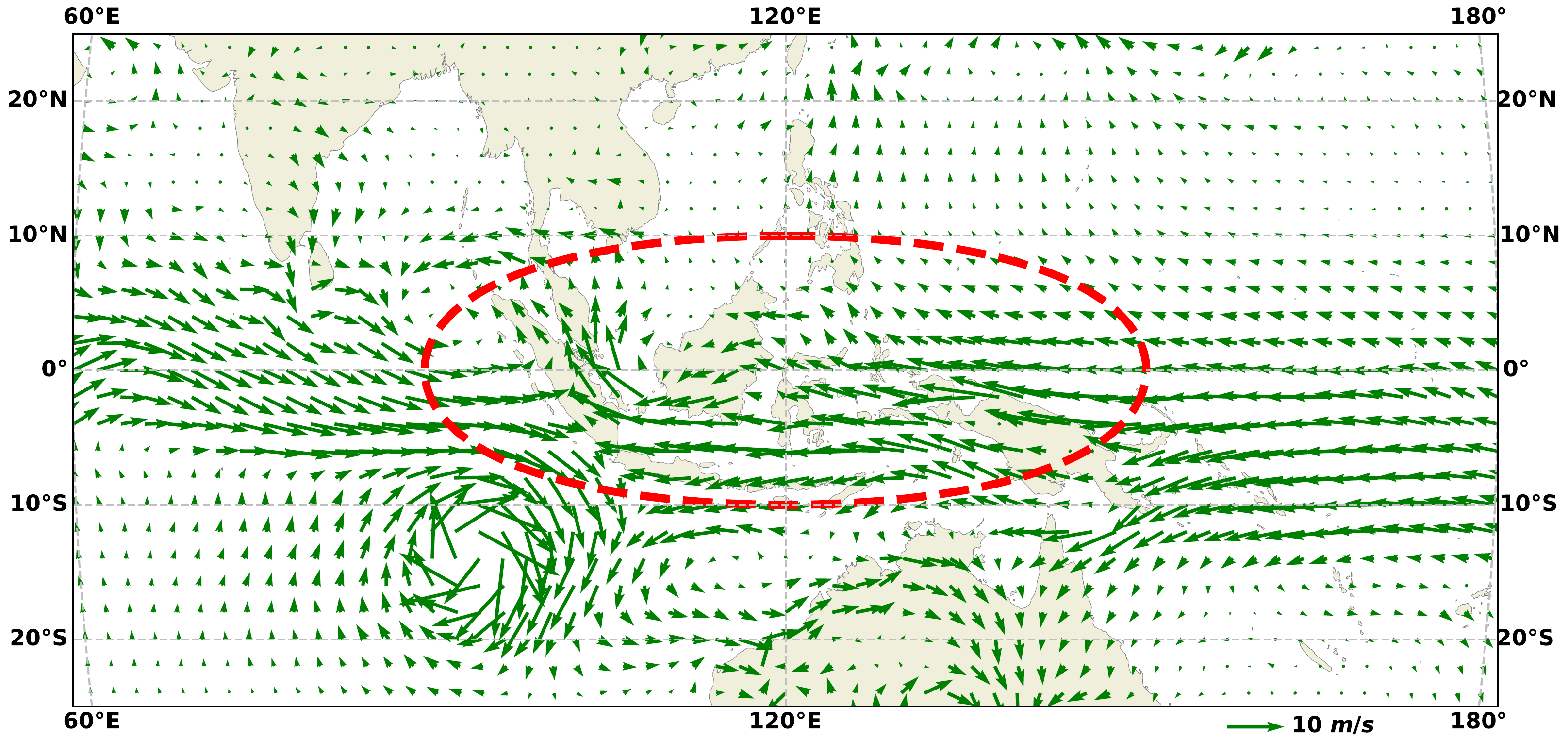}
\caption{850hPa anomaly wind vectors for the steady heating experiment after 20 days. The red dashed line outlines the region of steady heating.\label{fig:heating_850}}
\end{figure}

A closer examination of the response in the lower troposphere near the heating reveals a pattern similar to the classical Matsuno--Gill \cite{matsuno1966,gill1980} response to steady tropical heating (Fig.~\ref{fig:heating_850}). Along the equator, wind anomalies are convergent toward the western end of the heating region. This signature is associated with a Kelvin-wave response to the heating. Off the equator, the western end of the heating is flanked by cyclonic gyres in both hemispheres, which are associated with a mixed Rossby-gravity-wave response. Unlike idealized experiments, typically using the shallow-water equations, these solutions are influenced by surface boundary conditions, so that there are flow distortions over the Maritime Continent in particular, and myriad multiscale moist processes involving clouds and convection.

This experiment suggests that the Pangu-weather model responds qualitatively, if not quantitatively, consistent with idealized experiments for tropical heating. Anomalies emerge smoothly and locally from the heat source, and increases in amplitude with time as a nearly stationary wave response. Idealizing the problem further, to the zonal-mean DJF basic state produces a similar response, with a wave train extending across the North Pacific to North America (Fig.~S1), but with differences in phase and amplitude related to the basic state on which the waves propagate. The Southern Hemisphere response is also notably weaker for the zonal-mean state.

\subsection{Extratropical cyclone development\label{subsec:cyclone}}

\begin{figure}[h]
\noindent\includegraphics[height=.9\textheight]{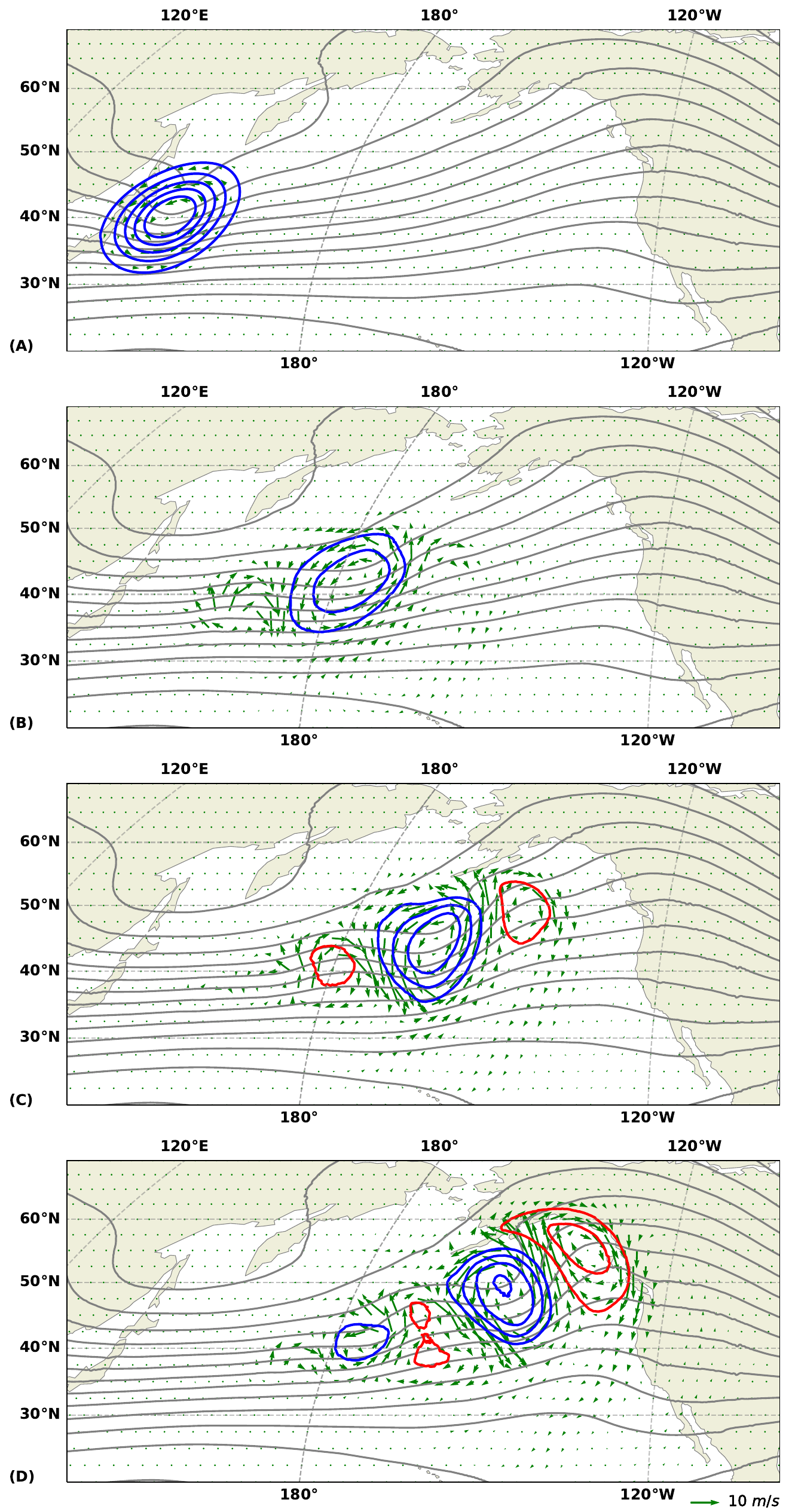}
\caption{Solution at 500hPa for a localized disturbance on the DJF atmosphere. The full geopotential height is shown by gray lines every 60m, and anomalies from the DJF average by red (positive) and blue (negative) lines every 20m; the zero contour is suppressed. Green arrows show the anomalous vector wind. Solutions are shown at (A) 0 days (the specified initial condition); (B) 2 days; (C) 3 days; and (D) 4 days.\label{fig:IVP_500} }
\end{figure}

\begin{figure}[h]
\noindent\includegraphics[height=.9\textheight]{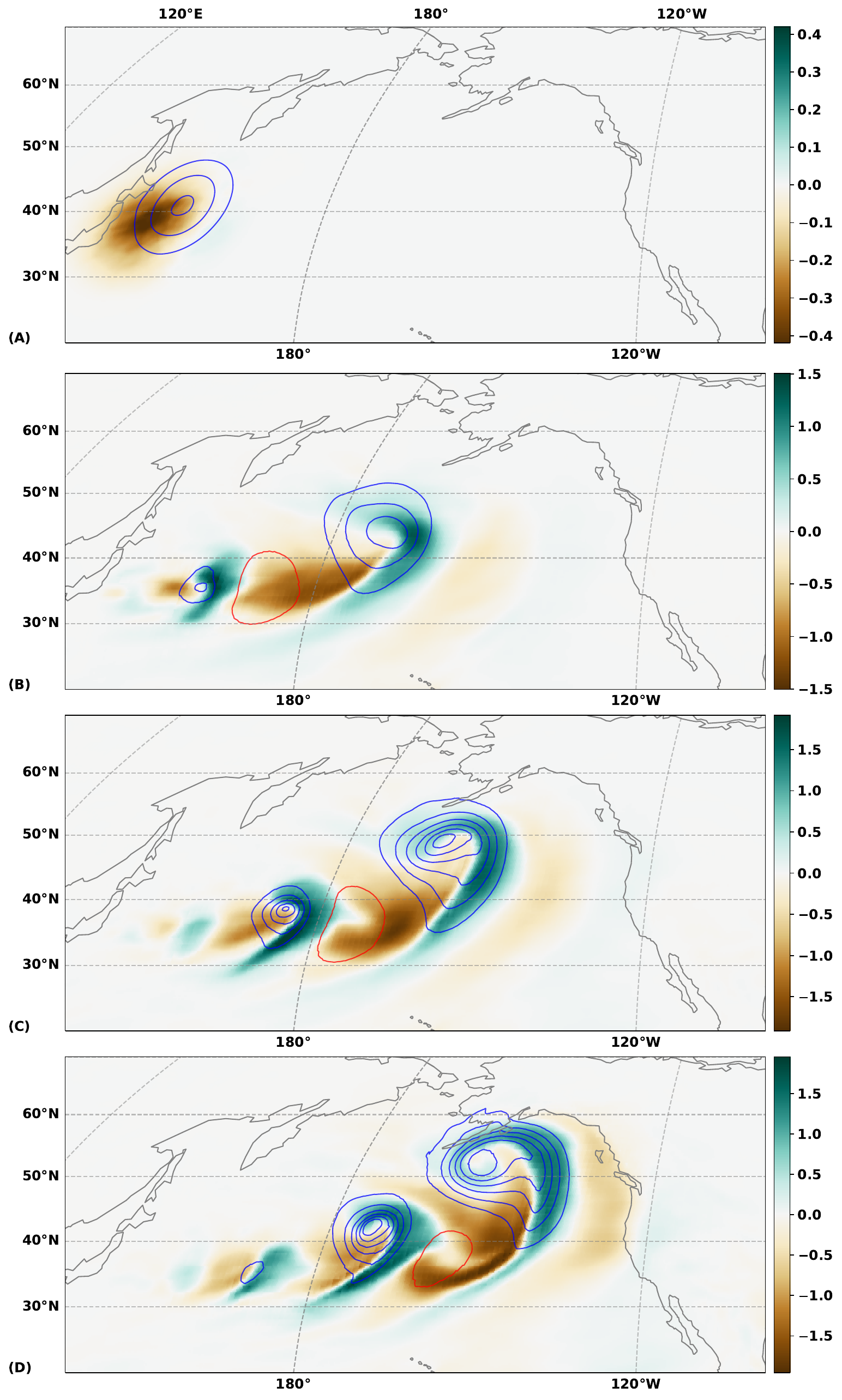}
\caption{Surface cyclones associated with the solution in Fig.~\ref{fig:IVP_500}. Anomalies in mean-sea-level pressure are shown every 2 hPa, with red (blue) lines for positive (negative) values; the zero contour is suppressed. Water vapor specific humidity anomalies (g/kg) at 850 hPa are shaded. Solutions are shown at (A) 0 days (the specified initial condition); (B) 2 days; (C) 3 days; and (D) 4 days.\label{fig:IVP_SFC}}
\end{figure}

The next experiment considers the time evolution of a localized 500hPa trough at the western end of the North Pacific storm track (Fig.~\ref{fig:IVP_500}A), which is the canonical initial condition preceding surface cyclogenesis \cite<e.g.,>{gyakum2000,hakim2003,yoshida2004}. After two days, the trough has progressed to the central Pacific, and begun to disperse, with the appearance of anticyclonic circulations both upstream and downstream (Fig.~\ref{fig:IVP_500}B). A surface cyclone develops to the east of the upper trough, with a smaller-scale secondary cyclone appearing upstream (Fig.~\ref{fig:IVP_SFC}B). By day 4, the upper trough has amplified and spread into a wave packet, with the leading edge along western North America (Fig.~\ref{fig:IVP_500}D), and a surface cyclone nearly coincident with the upper trough (Fig.~\ref{fig:IVP_SFC}D). Vertical alignment of extratropical cyclones is the hallmark of a developing cyclone that has reached the occluded phase of the life cycle. In contrast, the upstream surface cyclone remains downstream of the 500 hPa trough, and continues to deepen past day 4. A second upstream cyclone appears at day 4 west of the Dateline. These cyclones are accompanied by temperature anomalies having the largest horizontal gradients near the surface cold front (Fig.~S2).

All aspects of this idealized baroclinic development are consistent with observations and modeling \cite<e.g.,>{jablonowski2006} of the North Pacific storm track. In particular, disturbances at the upstream end of the storm track produce a baroclinic wave packet \cite{simmons1979}, which disperses and moves downstream at the group velocity (faster than the phase of individual troughs). As we find here, these solutions also show both upstream surface development and downstream upper-level development \cite{simmons1979,chang1993,hakim2003}. Moreover, the upstream surface development we observe here has relatively smaller spatial scale, resembling a ``polar low,'' which is frequently observed in winter over the North Pacific \cite<e.g.,>{mullen1983,rasmussen2003}. Curiously, these polar lows appear first at the surface, and have a warm core, suggestive of the importance of surface fluxes due to cold air moving over relatively warmer water \cite{emanuel1989}.

Idealizing the problem further, to the zonal-mean DJF atmosphere produces a similar response, with a wave packet that spreads downstream toward Europe by day 10 (Fig.~S3). Furthermore, repeating the experiment, but for summer conditions (JAS time mean) shows much weaker cyclone development, and an absence of polar lows (not shown). We conclude that Pangu-weather appears to have implicitly encoded the seasonally varying physical processes of oceanic extratropical cyclone development in the neural-network weights that govern the dynamical evolution of its prognostic variables.

\subsection{Geostrophic adjustment\label{subsec:geoadjust}}

Here we test an initial perturbation similar to the extratropical cyclone case, except that it is localized completely to the 500hPa field; it does not extend in the vertical, and every other field has zero anomaly. This type of initial condition is unbalanced since there is no wind or temperature anomalies, whereas outside the deep tropics one commonly finds the wind blowing along the height contours (as evident in Fig.~\ref{fig:IVP_500}A). This is a particularly hard test, and one that likely cannot be performed without additional modification using a physics-based model, since unbalanced initial conditions produce rapid oscillations that are difficult to resolve. Here we use the 1h, 3h, and 6h versions of the Pangu-weather models.

\begin{figure}
\noindent\includegraphics[height=.9\textheight]{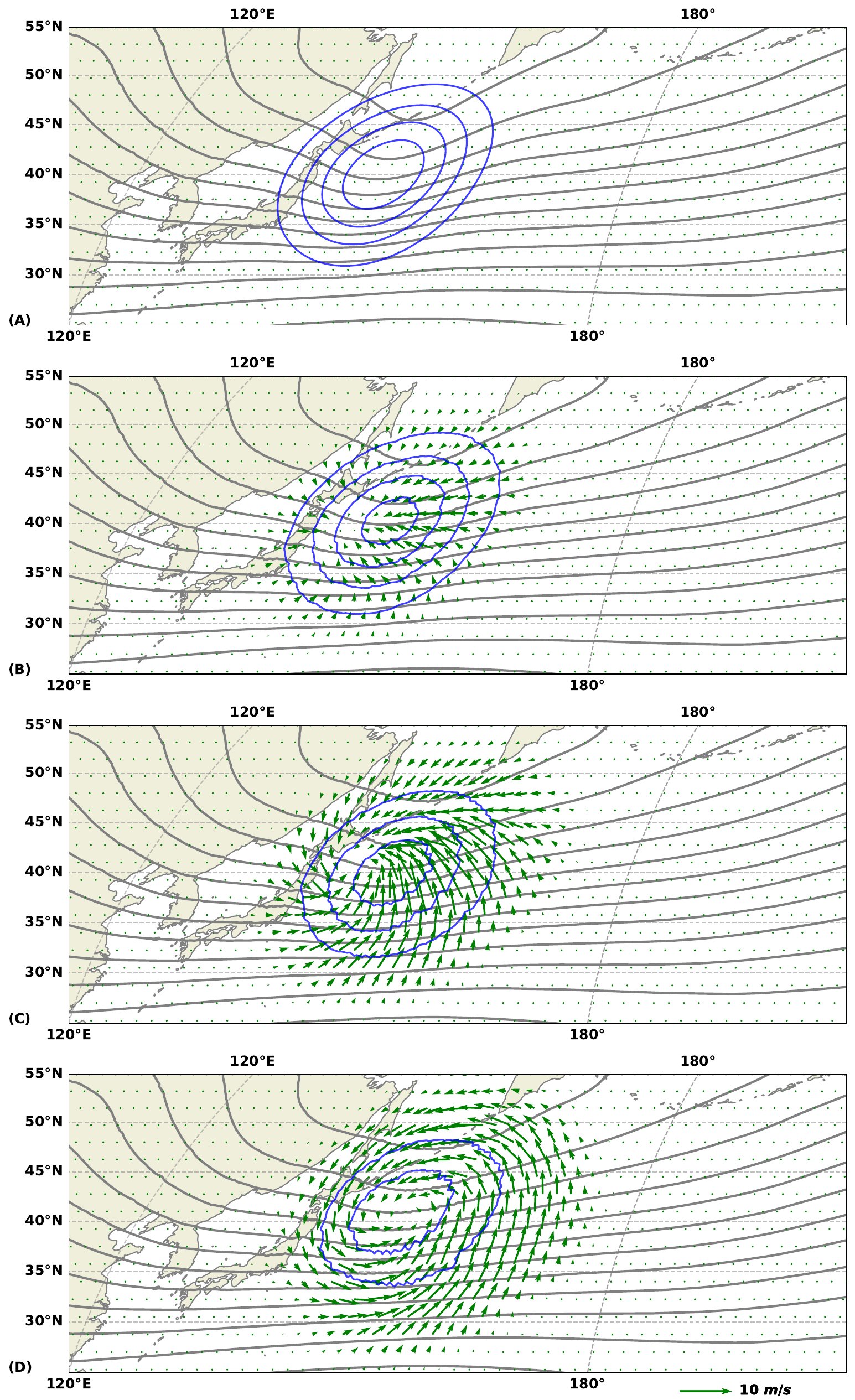}
\caption{Solution at 500hPa for the geostrophic adjustment problem consisting of a localized geopotential height disturbance on the DJF-averaged atmosphere. The full geopotential height is shown by gray lines every 60m, and negative anomalies by blue lines every 20m; the zero contour is suppressed. Green arrows show the anomalous vector wind. Solutions are shown at (A) t=0  (the specified initial condition); (B) 1 hour; (C) 3 hours; (D) 6 hours.\label{fig:geo_adjust}}
\end{figure}

At 1h, the wind accelerates from rest in the initial conditions to about 5 ms$^{-1}$, and is convergent on the area of low geopotential height (Fig.~\ref{fig:geo_adjust}A). The center of convergence is to the west of the lowest height, which increases to $-89$m from $-100$m in the initial condition. At 3h, the wind accelerates to a maximum of about 10 ms$^{-1}$, and remains convergent on the area of low height, for which the minimum has increased to $-74$m (Fig.~\ref{fig:geo_adjust}B). The wind direction has turned clockwise at all locations compared to the 1h solution, as one expects from Coriolis turning of the accelerating wind in the direction of the pressure gradient force. At 6h, the wind direction has continued to rotate clockwise such that it is nearly parallel to the geopotential height contours everywhere, reflecting a closer balance between the wind and geopotential height fields (Fig.~\ref{fig:geo_adjust}C). The height minimum has increased to $-58$m, reflecting a conversion of available potential energy to kinetic energy.

A quarter turn of a Foucault pendulum at 40$^{\circ}$ N takes $\sim$9 hours, so the adjustment in the wind field indicated by the Pangu-weather solution is consistent with physical expectations. Once again, we conclude that the solution for this idealized initial-value problem is qualitatively, if not quantitatively, consistent with the expected dynamics.

Repeating the experiment, except for an initial disturbance on the equator, produces a notably different response (Fig.~S4). The velocity field is again convergent on the area of low geopotential height, except in this case convergence is directed on the center of the low. The difference may be due to the basic-state jet stream in the previous case, with fast westerly winds and a strong meridional potential vorticity gradient that promotes westward Rossby-wave propagation. Another notable aspect of the equatorial case is slower Coriolis turning of the wind, and the fact that the model has learned about the asymmetry in this turning about the equator. An analysis of the time difference of the anomalous zonal wind on the equator reveals signals that propagate in both directions at around 20ms$^{-1}$ (Fig.~S5), typical of tropical gravity waves. A weaker signal is also evident at the speed of sound (dashed black lines). Finally, we note that this figure shows the incompatibility between the different versions of the Pangu-weather model, with abrupt differences in the time tendency at intervals of 3, 6, and 24 hours. These single-step ``shocks'' do not appear to adversely affect the solution at subsequent times, but will affect temporal diagnostic calculations that span several time steps of the model.

\subsection{Atlantic hurricane development\label{sec:hurricane}}

The last example concerns the evolution of a localized disturbance in the subtropics for the July--September (JAS) averaged conditions. Seeds of Atlantic hurricanes take the form of weak low pressure systems, which may develop into mature storms given the right environmental conditions. Finite-amplitude disturbances are thought to be needed to reduce the time to development while the storm is in a favorable environment \cite<e.g.,>{mcbride1981,nolan2007}. Here we perform experiments for a localized area of low pressure at a reference location (15$^{\circ}$N, 40$^{\circ}$W), and vary the initial amplitude. The three-dimensional perturbation is constructed similarly to the initial condition for the extratropical cyclone case, by regressing all variables and locations onto the mean-sea-level pressure at the reference location.

\begin{figure}
\noindent\includegraphics[width=.9\textwidth]{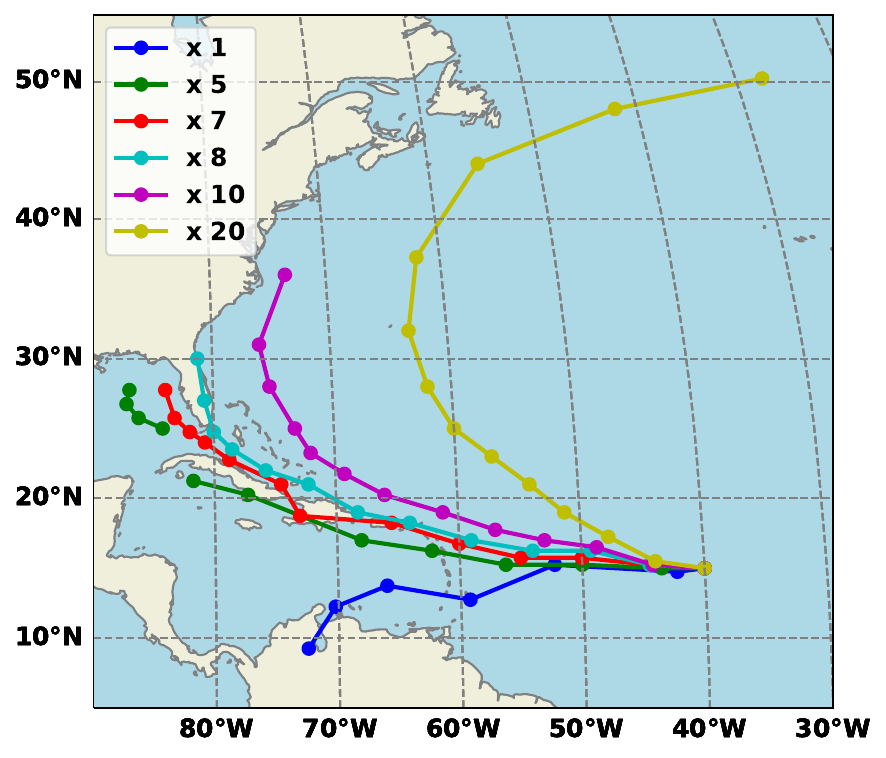}
\caption{Tracks of mean-sea-level pressure minima for experiments seeding Atlantic hurricanes on the July--September-averaged atmosphere. All experiments are initialized with a surface-based low-pressure system at 15$^{\circ}$N, $40^{\circ}$W, and initial amplitude by a scaling factor on the climatological JAS standard deviation at that location (indicated in the legend).\label{fig:hurricane_map}}
\end{figure}

Results show that the low pressure systems take a familiar track toward the northwest around the climatological subtropical area of high pressure (Fig.~\ref{fig:hurricane_map}). Stronger initial conditions take a progressively northward track, which is consistent with the known physical basis due to increasing amplitude of azimuthal wavenumber-one asymmetries (``$\beta$ gyres''). Although Pangu-weather may at best poorly resolve these features, the weights in the neural network have identified this physical relationship between the strength of tropical cyclones and a northward track. 

\begin{figure}
\noindent\includegraphics[width=.9\textwidth]{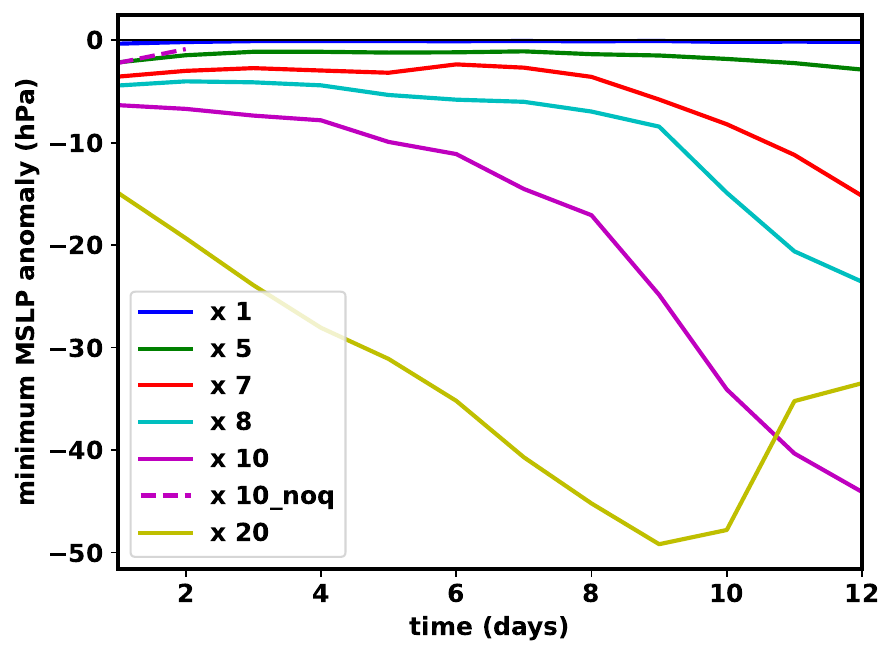}
\caption{Intensity of the low pressure systems tracked in Fig.~\ref{fig:hurricane_map} in terms of anomalous mean-sea-level pressure (hPa) as a function of time (days).\label{fig:hurricane_intensity}}
\end{figure}

For initial disturbances with anomalous mean-sea-level pressure less than about $\sim$5hPa, the storms do not intensify, whereas initial disturbances stronger than this rapidly intensify (Fig.~\ref{fig:hurricane_intensity}). An additional experiment for the 10x disturbance was performed by setting the water vapor specific humidity to zero and, unlike the original case that rapidly develops, the dry system rapidly decays. Pangu-weather doesn't explicitly model condensational heating, but the model weights have the conditional association between water vapor content and the development of tropical cyclones.

\section{Conclusions\label{sec:conclusions}}

We have tested the Pangu-weather deep-learning weather prediction model on a set of four canonical experiments aimed at probing its dynamical response to local perturbations. These perturbations are helpful for determining whether disturbances evolve and propagate in a physically meaningful way. Our hypothesis at the outset of this work was that these localized features would immediately produce a global response because there is no constraint to prevent this during model training. The fact that every experiment produces signal propagation and structural evolution qualitatively in accord with previous research suggests that the model has encoded realistic physics. While we do not make a direct comparison to solutions from a physics-based model, the results here provide proof-of-concept motivating such experiments. We note that, due to differences in numerics and parameterizations for unresolved scales and processes, solutions from physics-based models for these experiments will differ in details, and it would be interesting to see if the Pangu-weather solutions fall within the uncertainty of the physics-based models.

Results from the canonical experiments show qualitative, if not quantitative, agreement with studies of similar phenomena in the literature. This agreement ranges from hourly timescales for the geostrophic adjustment process to approximately steady features beyond 10~days associated with stationary tropical heating. Highlights from these experiments include: a Matsuno--Gill response and extratropical planetary wave response to steady tropical heating; baroclinic wave-packet emergence and polar low development in the cold-air mass associated with a North Pacific extratropical cyclone; divergent flow yielding to rotational flow for an unbalanced initial condition; and the importance of initial-vortex amplitude and water vapor in the development and track of Atlantic hurricanes.

We conclude that the Pangu-weather model encodes realistic physics for the experiments considered here, motivating future basic research using this tool. Several attributes make this model particularly powerful for atmospheric dynamics and scientific hypothesis testing. First, the simulations are computationally inexpensive compared to traditional global weather models. This enables large ensembles, including iterations over varying parameters, initial conditions, and perturbations to model output. Second, experiments are extremely easy to configure, and the model is very forgiving in aspects that physics models are not. For example, initial imbalances in physics-based models can produce spurious oscillations at the model time step that are difficult to remove or filter without affecting the resolved scales of interest. Therefore, we speculate that models like Pangu-weather might be particularly useful for rapid evaluation of hypotheses, allowing tests over a wide range of ideas to quickly narrow the scope of investigation for experiments using expensive physics-based models. Among many possibilities, one particularly interesting path of research employs deep-learning models to examine multiscale phenomena involving convective clouds, such as the Madden-Julian Oscillation, where physics-based models and theory have not yet approximated the essential physical processes.

\section*{Open Research Section}
All code and data access will be released in an open Github repository upon acceptance of the paper for publication.

%This section MUST contain a statement that describes where the data supporting the conclusions can be obtained. Data cannot be listed as ''Available from authors'' or stored solely in supporting information. Citations to archived data should be included in your reference list. Wiley will publish it as a separate section on the paper’s page. Examples and complete information are here:
%https://www.agu.org/Publish with AGU/Publish/Author Resources/Data for Authors

\acknowledgments
We thank Steve Penny for conversations related to deep-learning models in the geosciences, and Mike Pritchard for comments on an earlier draft of the manuscript.

%%%%%%%%%%%%%%%%%%%%%%%%%%%%%%%%%%%%%%%%%%%%%%%
% REFERENCES and BIBLIOGRAPHY
%
% \bibliography{<name of your .bib file>} don't specify the file extension
% don't specify bibliographystyle
%
%%%%%%%%%%%%%%%%%%%%%%%%%%%%%%%%%%%%%%%%%%%%%%%

\bibliography{hakim_masanam_refs.bib}

%Reference citation instructions and examples:
%
% Please use ONLY \cite and \citeA for reference citations.
% \cite for parenthetical references
% ...as shown in recent studies (Simpson et al., 2019)
% \citeA for in-text citations
% ...Simpson et al. (2019) have shown...
%
%
%...as shown by \citeA{jskilby}.
%...as shown by \citeA{lewin76}, \citeA{carson86}, \citeA{bartoldy02}, and \citeA{rinaldi03}.
%...has been shown \cite{jskilbye}.
%...has been shown \cite{lewin76,carson86,bartoldy02,rinaldi03}.
%... \cite <i.e.>[]{lewin76,carson86,bartoldy02,rinaldi03}.
%...has been shown by \cite <e.g.,>[and others]{lewin76}.
%
% apacite uses < > for prenotes and [ ] for postnotes
% DO NOT use other cite commands (e.g., \citet, \citep, \citeyear, \nocite, \citealp, etc.).
%

\end{document}